\documentclass[journal]{IEEEtran}

\usepackage{epsfig}
\usepackage{subfigure}
\usepackage{amssymb}
\usepackage{amstext}
\usepackage{amsmath}
\usepackage{amsthm}
\usepackage{multicol}
\usepackage{pslatex}
\usepackage{float}
\usepackage[small]{caption}
\usepackage[update,prepend]{epstopdf}
\usepackage{lmodern}
\usepackage{cite}
\usepackage{listings}

\def\postbreak{\raisebox{0ex}[0ex][0ex]{\ensuremath{\hookrightarrow\space}}}
\begin{document}

\title{Limitations of PLL simulation: \\ hidden oscillations in MatLab and SPICE
}

\author{\IEEEauthorblockN{
  Bianchi G.\IEEEauthorrefmark{1},
  Kuznetsov N.~V.\IEEEauthorrefmark{2}\IEEEauthorrefmark{3},
  Leonov G.~A.\IEEEauthorrefmark{2},
  Yuldashev M.~V.\IEEEauthorrefmark{2},
  Yuldashev R.~V.\IEEEauthorrefmark{2}
}
\\
\IEEEauthorblockA{
\IEEEauthorrefmark{1}
Advantest Europe GmbH}
\\
\IEEEauthorblockA{\IEEEauthorrefmark{2}
Faculty of Mathematics and Mechanics,\
Saint-Petersburg State University, Russia} \\
\IEEEauthorblockA{
\IEEEauthorrefmark{3}
  Dept. of Mathematical Information Technology,\
  University of Jyv\"{a}skyl\"{a}, Finland \\ email: nkuznetsov239@gmail.com
}
\thanks{
  Accepted to IEEE 7th International Congress on Ultra Modern Telecommunications and Control Systems, 2015
}
}

\maketitle

\begin{abstract}                
 Nonlinear analysis of the phase-locked loop (PLL) based circuits
 is a challenging task, thus
 in modern engineering literature
 simplified mathematical models and simulation are widely used for their study.
 In this work the limitations of numerical approach is discussed
 and it is shown that, e.g. hidden oscillations may not be found by simulation.
 Corresponding examples in SPICE and MatLab,
 which may lead to wrong conclusions concerning the operability of PLL-based circuits,
 are presented.
\end{abstract}

\IEEEpeerreviewmaketitle

\section{Introduction}

The phase-locked loop based circuits (PLL)
are widely used nowadays in various applications.
PLL is essentially a nonlinear control system
and its rigorous analytical analysis is a challenging task.
Thus, in practice, simulation is widely used
for the study of PLL-based circuits
(see, e.g.
\cite{Bianchi-2005-book,Best-2007,Tranter-2010-book,Talbot-2012-book}).
At the same time,
simulation of nonlinear control system may lead to wrong conclusions,
e.g. recent work \cite{LauvdalMF-1997} notes that
\emph{stability in simulations may not imply
stability of the physical control system, thus
stronger theoretical understanding is required}.

In this work the two-phase PLL is studied
and corresponding examples,
where simulation leads to unreliable results,
is demonstrated in SPICE and MatLab.

\section{PLL operation}
Typical analog PLL consists of the following elements:
a voltage-controlled oscillator (VCO),
a linear low-pass filter (LPF),
a reference oscillator (REF),
and an analog multiplier $\otimes$ used as the phase detector (PD).
The phase detector compares the phase of VCO signal
 against the phase of reference signal;
the output of the PD (error voltage) is proportional
 to the phase difference between its two inputs.
Then the error voltage is filtered by the loop filter (LPF).
The output of the filter is fed to the control input of the VCO,
which adjusts the frequency and phase to synchronize with the reference signal.

\begin{figure}[H]
\centering
\includegraphics[scale=0.4]{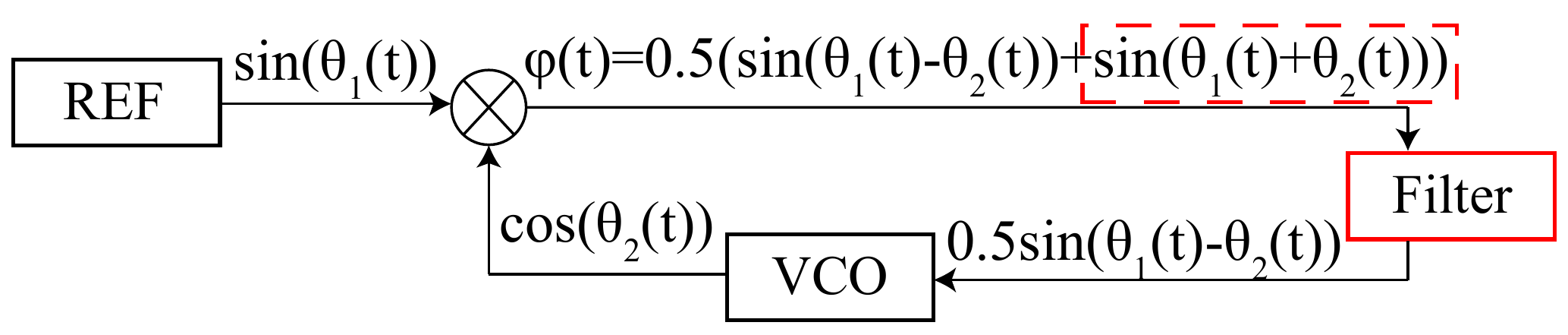}
\caption{Operation of classical phase-locked loop for sinusoidal signals}
\label{pll_expl_sin}
\end{figure}

Consider a signal space model of the classical analog PLL
with a multiplier as a phase detector (see Fig.~\ref{pll_expl_sin}).

Suppose that both waveforms of VCO
and the reference oscillator signals are sinusoidal
\footnote{\
  Other waveforms can be similarly considered \cite{LeonovKYY-2012-TCASII,LeonovKYY-2015-SIGPRO,KuznetsovLSYY-2015-PD}}.
(see Fig.~\ref{pll_expl_sin}).
The low-pass filter passes low-frequency signal
 $\sin(\theta_1(t)-\theta_2(t))$ and attenuates high-frequency signal
 $\sin(\theta_1(t)+\theta_2(t))$.

The averaging under certain conditions \cite{KrylovB-1947,KudrewiczW-2007,LeonovKYY-2012-TCASII,LeonovK-2014,LeonovKYY-2015-SIGPRO}
and approximation
$\varphi(t) \approx \sin(\theta_1(t)-\theta_2(t))$
allow one to proceed from the analysis of the signal space model
to the study of PLL model in the signal's phase space.
Rigorous consideration of this point is often omitted
(see, e.g. classical books \cite[p.12,p15-17]{Viterbi-1966}, \cite[p.7]{Gardner-1966})
while it may lead to unreliable results
(see, e.g. \cite{KuznetsovKLNYY-2015-ISCAS,BestKKLYY-2015-ACC}).

One of the approaches to avoid this problem is the use of two-phase modifications of PLL,
which does not have high-frequency oscillations at the output
of the phase detector \cite{Emura-1982}.

\section{Two-phase PLL}
\begin{figure}[H]
 \centering
 \includegraphics[scale=0.45]{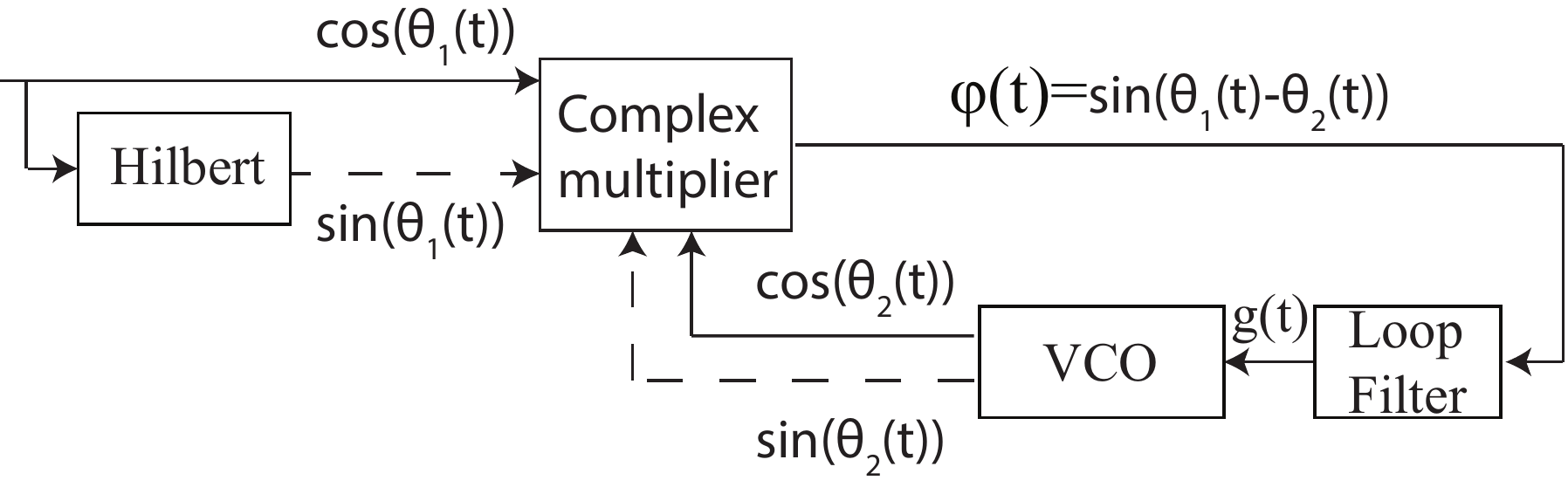}
 \caption{Two-phase PLL}
 \label{fig:hilbert_pll}
\end{figure}

\begin{figure}[h]
 \centering
 \includegraphics[scale=0.7]{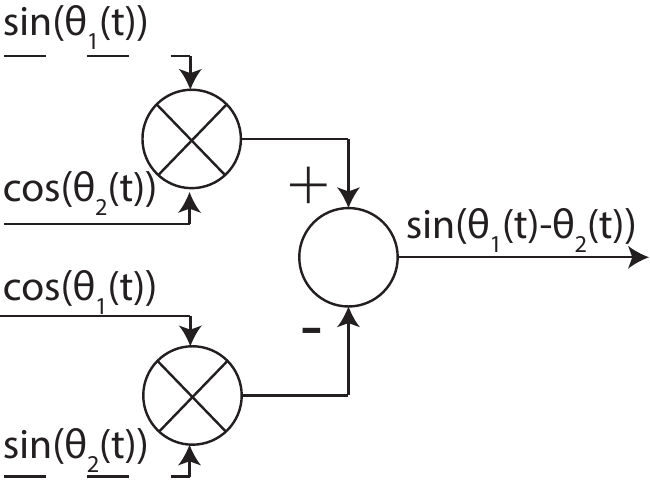}
 \caption{Phase detector in two-phase PLL }
 \label{funky_pd}
\end{figure}

Consider two-phase PLL model in Fig.~\ref{fig:hilbert_pll}.
Here a carrier is $\cos(\theta_1(t))$  with $\theta_1(t)$ as a phase
and the output of Hilbert block is $\sin(\theta_1(t))$.
The VCO generates oscillations $\sin(\theta_2(t))$ and $\cos(\theta_2(t))$
with $\theta_2(t)$ as a phase.
Fig.~\ref{funky_pd} shows the structure of phase detector (complex multiplier).
The phase detector consists of two analog multipliers and analog subtractor.
The output of PD is
$
   \varphi(t) = \sin(\theta_1(t))\cos(\theta_2(t)) - \cos(\theta_1(t))\sin(\theta_2(t)) =
   \sin(\theta_1(t)-\theta_2(t))
$
In this case there is no high-frequency component
at the output of phase detector.
It is reasonable to introduce a phase detector gain $\frac{1}{2}$
(e.g. to consider additional loop filter gain equal to $0.5$)
to make it the same as a classic PLL phase detector characteristic:
\begin{equation}
\label{phi}
\begin{aligned}
  & \varphi(t) = \frac{1}{2} \sin(\theta_1(t)-\theta_2(t)).
\end{aligned}
\end{equation}

Consider a loop filter with the transfer function $H(s)$.
The relation between input $\varphi(t)$ and output $g(t)$
of the loop filter is as follows
\begin{equation}
\label{filter-eq}
\begin{aligned}
& \dot x = Ax + b\varphi(t), \quad g(t) = c^*x + h\varphi(t),
\\
& H(s) = c^*(A - sI)^{-1}b - h.
\end{aligned}
\end{equation}

The control signal $g(t)$ is used to adjust the VCO phase to
the phase of the input carrier signal:
\begin{equation}
\label{vco-eq}
\begin{aligned}
& \theta_2(t)
  = \int_0^{t}\omega_2(\tau)d\tau
  = \omega_{free}t + L\int_0^{t}g(\tau)d\tau,
\end{aligned}
\end{equation}
where $\omega_2^{\text{free}}$ is the VCO free-running frequency
(i.e. for $g(t)\equiv 0$) and $L$ the VCO gain.

Next examples show the importance of analytical methods
for investigation of PLL stability.
It is shown that the use of default simulation parameters
for the study of two-phase PLL in MatLab and SIMULINK can lead to
wrong conclusions concerning the operability of the loop,
e.g. the pull-in (or capture) range
(see discussion of rigorous definitions in \cite{KuznetsovLYY-2015-IFAC-Ranges,LeonovKYY-2015-TCAS}).

\begin{figure*}[]
\centering
 \includegraphics[width=0.8\textwidth]{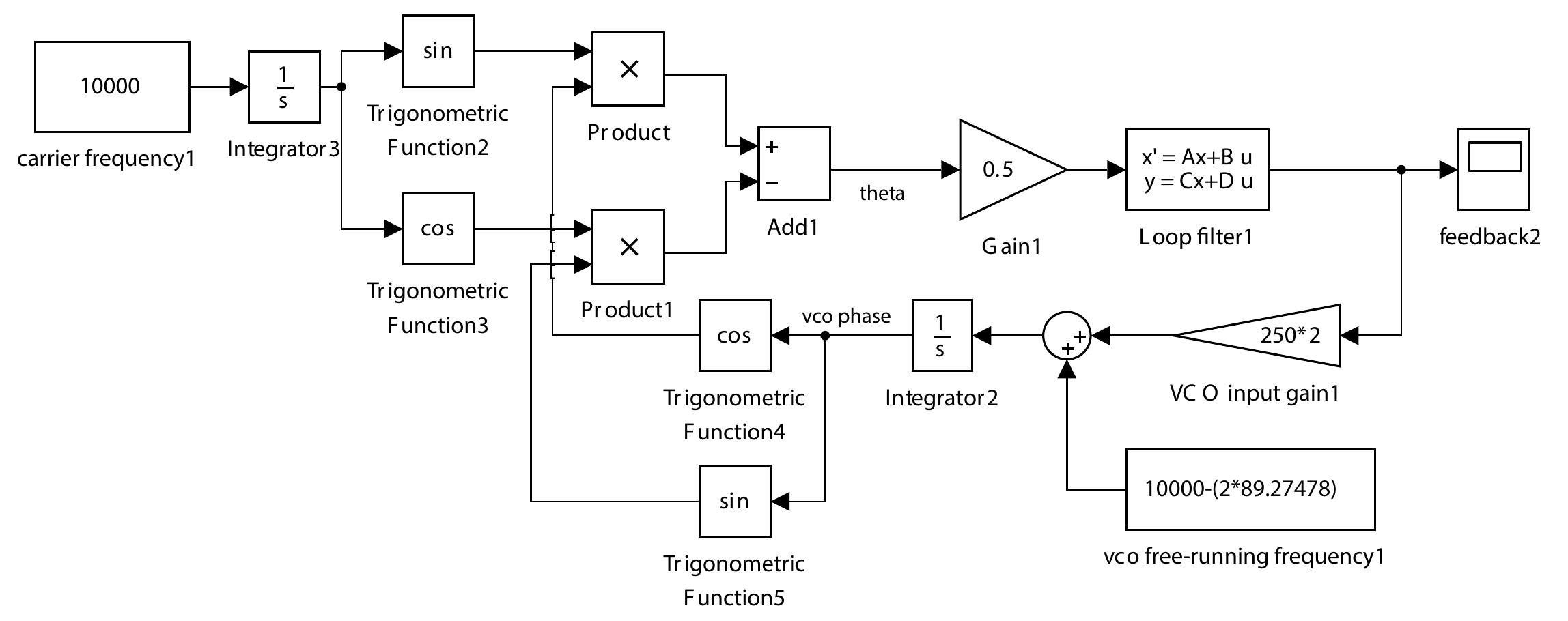}
 \caption{Model of two-phase PLL in MatLab Simulink}
 \label{simulink-2phase-pll-model}
\end{figure*}

\section{Simulation in MatLab}
Consider a passive lead-lag loop filter
with the transfer function
$H(s) = \frac{1+s \tau_2}{1+s(\tau_1 + \tau_2)}$,
$\tau_1 = 0.0448$, $\tau_2 = 0.0185$
and the corresponding parameters
$A = -\frac{1}{\tau_1+\tau_2}$,
$b = 1 - \frac{\tau_2}{\tau_1+\tau_2}$,
$c = \frac{1}{\tau_1+\tau_2}$,
$h = \frac{\tau_2}{\tau_1+\tau_2}$.
The model of two-phase PLL in MatLab is shown in Fig.~\ref{simulink-2phase-pll-model}
(see more detailed description of simulating PLL based circuits in MatLab Simulink
in
\cite{KuznetsovKLSYY-2014-ICUMT-BPSK,KuznetsovKLNYY-2014-ICUMT-QPSK,KuznetsovLNSYY-2012-IEEE-PLL}).

For the case of the passive lead-lag filter
a recent work \cite[p.123]{Margaris-2004} notes that
``{\it the determination of the width of the capture range together
with the interpretation of the capture effect in the second order type-I loops
have always been an attractive theoretical problem.
This problem has not yet been provided with a satisfactory solution}''.
Below we demonstrate that in this case a numerical simulation may give wrong estimates
and should be used very carefully.

In Fig.~\ref{simulink-2phase-pll-model} we use the block \emph{Loop filter}
to take into account the initial filter state $x(0)$;
the initial phase error $\theta_\Delta(0)$
can be taken into account by the property \emph{initial data} of the \emph{Intergator} blocks\footnote{
  Following the classical consideration \cite[p.17, eq.2.20]{Viterbi-1966}\cite[p.41, eq.4-26]{Gardner-1966},
  where the filter's initial data is omitted,
  the filter is often represented in MatLab Simulink as the block \emph{Transfer Fcn}
  with zero initial state
  (see, e.g. \cite{BrigatiFMPP-2001,NicolleTMOJ-2007,Zucchelli-2007,KoivoE-2009,KaaldLHS-2009}).
  It is also related to the fact that the transfer function (from $\varphi$ to $g$)
  of system \eqref{filter-eq}
  is defined by the Laplace transformation for zero initial data $x(0) \equiv 0$.
}.
Note that the corresponding initial states in SPICE (e.g. capacitor's initial charge  and inductor's initial currents)
are zero by default but can be changed manually.

\begin{figure}[h]
  \centering
  \includegraphics[width=0.8\linewidth]{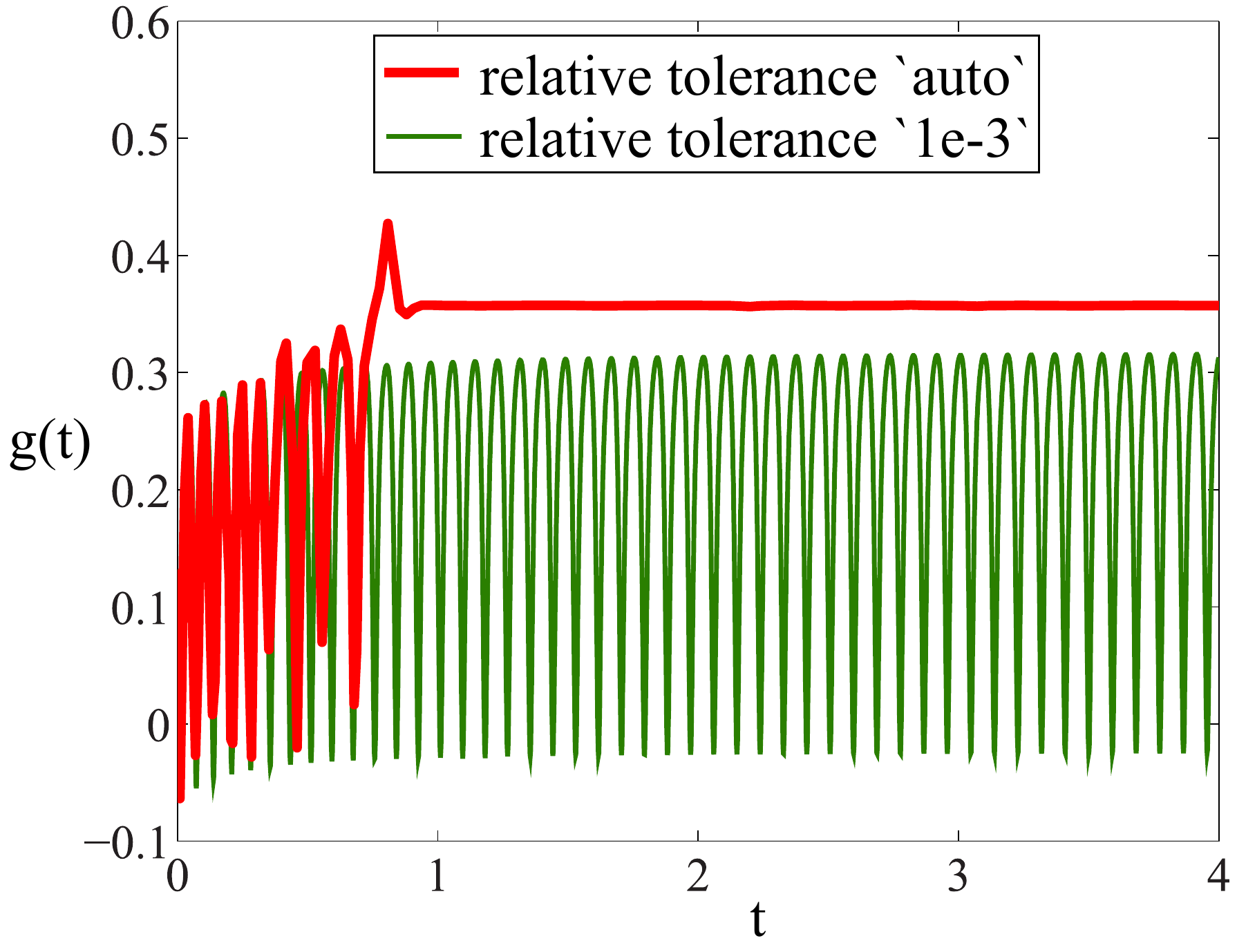}
  \caption{
  Simulation of two-phase PLL.
  Filter output $g(t)$ for the initial data $x_0\!=\!0.1318, \theta_{\Delta}(0)\!=\!0$
  obtained for default ``auto'' relative tolerance (red) --- acquires lock,
  relative tolerance set to ``1e-3''(green) --- does not acquire lock.}
  \label{pll_hidden}
\end{figure}

In Fig.~\ref{pll_hidden}
the two-phase PLL model simulated with relative tolerance set to ``1e-3''
or smaller
does not acquire lock (black color),
but the PLL model in signal's phase space simulated in MatLab Simulink
with standard parameters (a relative tolerance set to ``auto'')
 acquires lock (red color).
Here the input signal frequency is $10000$, the VCO free-running frequency
 $\omega_2^{\text{free}}= 10000 - 178.9$,
the VCO input gain is $L=500$,
 the initial state of loop filter is $x_0 = 0.1318$
 \footnote{Almost each initial state from the interval $[1,2]$ gives similar results.},
and the initial phase difference is $\theta_{\Delta}(0) = 0$.

\section{Simulation in SPICE}
In this section the previous example is reconstructed in SIMetrix,
which is one of the commercial versions of SPICE.

\begin{figure*}
\centering
 \includegraphics[width=0.8\textwidth]{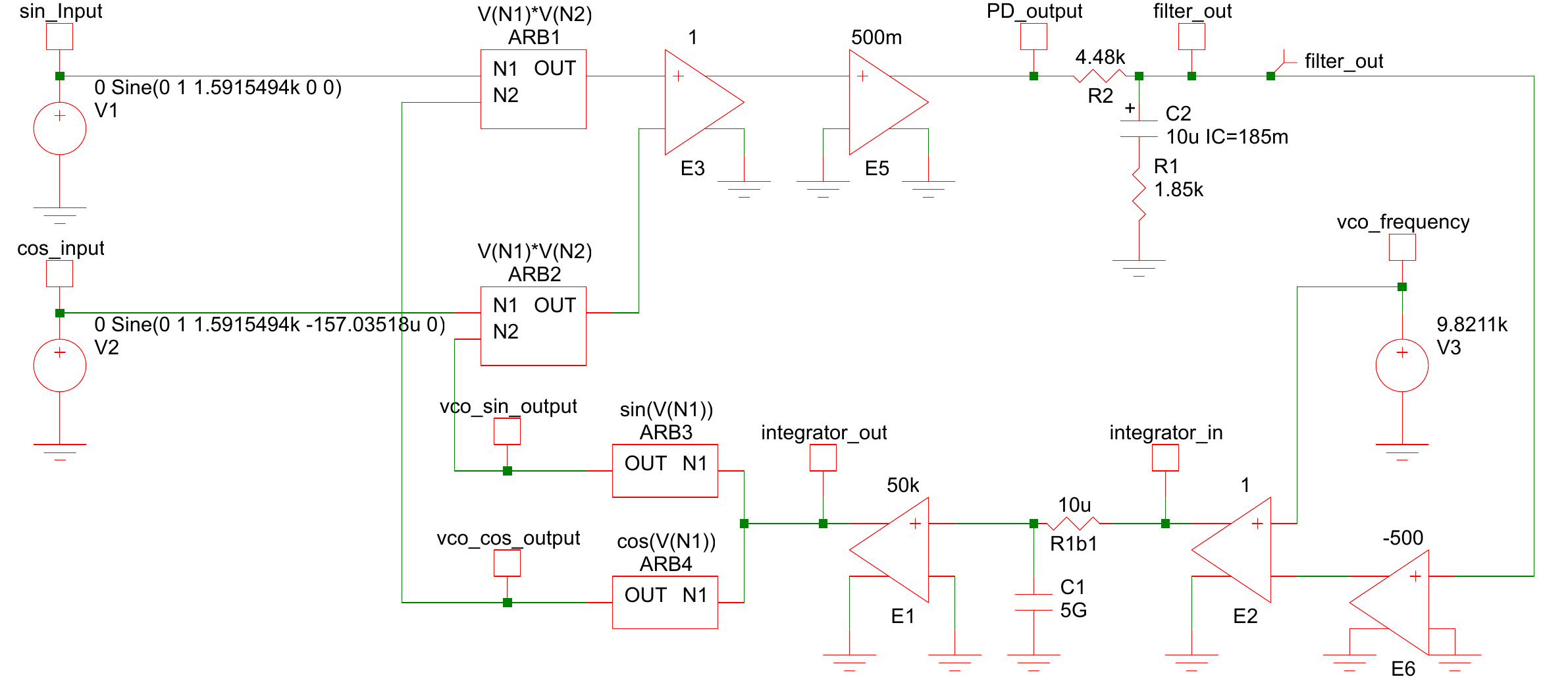}
 \caption{SPICE model of two-phase PLL in SIMetrix}
 \label{simetrix-pll}
\end{figure*}

Consider SIMetrix model of two-phase PLL shown
in Fig.~\ref{simetrix-pll}.
The input signal and the output of Hilbert block in Fig.~\ref{fig:hilbert_pll}
are modeled by sinusoidal voltage sources V1 (a frequency parameter is $1.5915494k$)
and V2 (a frequency parameter is $1.5915494k$ and a phase is $90$)
(sin\_input and cos\_input).
A complex multiplier in Fig.~\ref{funky_pd} is modeled
as two arbitrary sources ARB1 and ARB2 with definitions set to $V(N1)^*V(N2)$.
To subtract the output signals of multipliers,
Voltage Controlled Voltage Source (E3) is used.
Phase detector gain (E5) is equal to $\frac{1}{2}$.
Loop Filter in Fig.~\ref{fig:hilbert_pll} is modeled as a passive lead-lag
filter with resistor R2, capacitor C2, and resistor R1.
The input gain of VCO (E6) is equal to $-500$.
VCO self frequency\footnote{Zero input response (ZIR) frequency}
 (DC Voltage Source V3) is set to $9.8211k$.
Voltage Controlled Voltage Source E2 summarizes a VCO self frequency
and a control signal from E6.
Resistor R1b1 ($10u$), capacitor C1 ($5G$), and amplifier E1($50k$)
form an integrator\footnote{The same results could qualitatively be obtained using $10K$ for the resistor \
 and $5$ Farad for the capacitor (keeping $RC$ constant)}.
The VCO waveforms are defined by arbitrary blocks ARB3 (with
the function $\sin(V(N1))$)
 and ARB4 (with the function $\cos(V(N1))$)).
Netlist for the model, generated by SIMetrix,
is as follows:
\begin{lstlisting}
*#SIMETRIX
V1 sin_Input 0 0 Sine(0 1 1.5915494k 0 0)
V2 cos_input 0 0 Sine(0 1 1.5915494k -157.03518u 0)
V3 vco_frequency 0 9.8211k
R1 C2_N 0 1.85k
R2 filter_out PD_output 4.48k
X$ARB1 sin_Input vco_cos_output ARB1_OUT $$arbsourceARB1 pinnames: N1 N2 OUT
.subckt $$arbsourceARB1 N1 N2 OUT
B1 OUT 0 V=V(N1)*V(N2)
.ends
X$ARB2 cos_input vco_sin_output E3_CN $$arbsourceARB2 pinnames: N1 N2 OUT
.subckt $$arbsourceARB2 N1 N2 OUT
B1 OUT 0 V=V(N1)*V(N2)
.ends
X$ARB3 integrator_out vco_sin_output $$arbsourceARB3 pinnames: N1 OUT
.subckt $$arbsourceARB3 N1 OUT
B1 OUT 0 V=sin(V(N1))
.ends
X$ARB4 integrator_out vco_cos_output $$arbsourceARB4 pinnames: N1 OUT
.subckt $$arbsourceARB4 N1 OUT
B1 OUT 0 V=cos(V(N1))
.ends
E1 integrator_out 0 E1_CP 0 50k
E2 integrator_in 0 vco_frequency E2_CN 1
C1 E1_CP 0 5G
C2 filter_out C2_N 10u IC=185m  BRANCH={IF(ANALYSIS=2,1,0)}
E3 E3_P 0 ARB1_OUT E3_CN 1
E5 PD_output 0 E3_P 0 500m
E6 E2_CN 0 filter_out 0 -500
R1b1 integrator_in E1_CP 10u
.GRAPH filter_out curveLabel= filter_out nowarn=true ylog=auto xlog=auto disabled=false
.TRAN 0 5 0 1m UIC
.OPTIONS minTimeStep=1m
+  tnom=27
\end{lstlisting}

\begin{figure}[h]
\centering
 \includegraphics[width=0.45\textwidth]{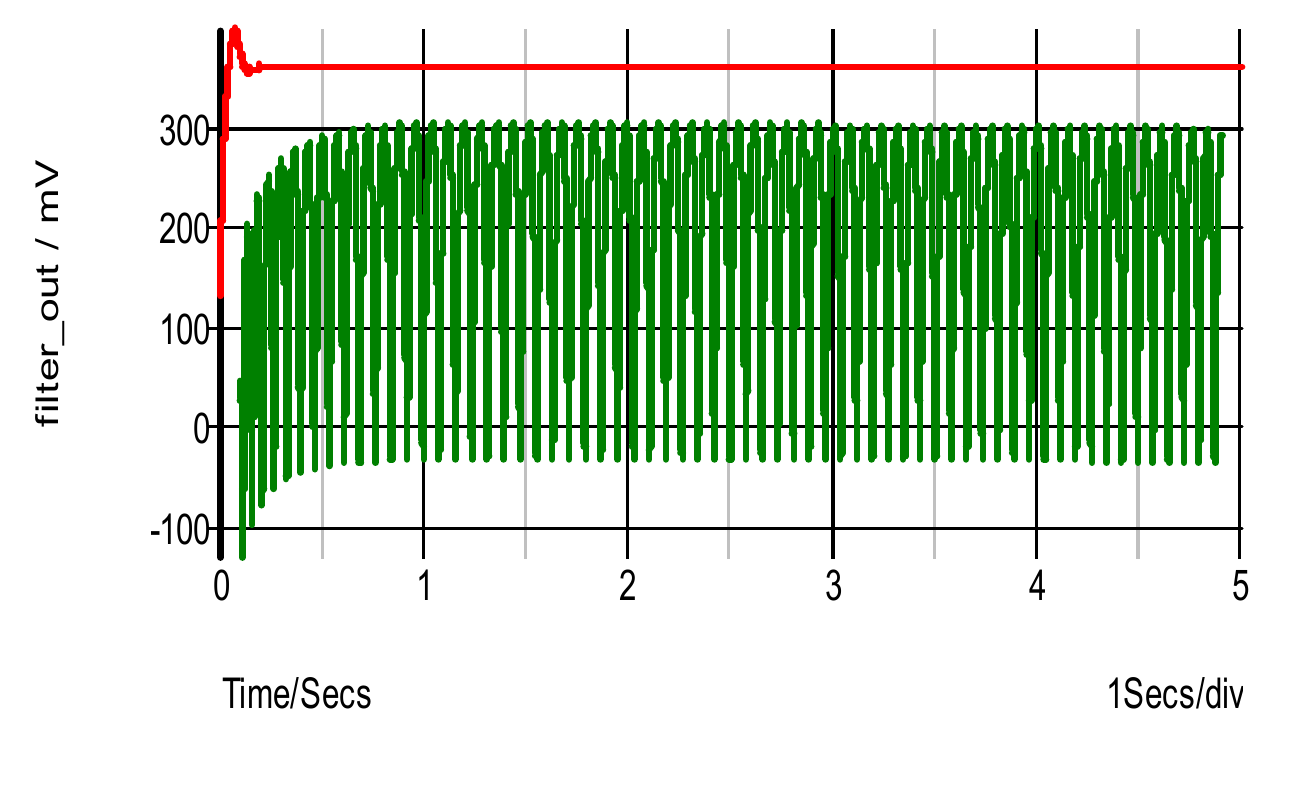}
 \caption{Hidden oscillations in SPICE}
 \label{simetrix-hidden}
\end{figure}

In Fig.~\ref{simetrix-hidden} are shown simulation results
in SPICE, which are close to the simulation results in MatLab Simulink
(see Fig.~\ref{pll_hidden}).
For default simulation parameters in SIMetrix
two-phase PLL synchronizes to the reference signal (red line).
However, if we choose smaller simulation step ($1m$),
the simulation reveals an oscillation (green line).
\section{Mathematical reasoning}
Two-phase PLL is described by \eqref{phi}, \eqref{vco-eq},
and \eqref{filter-eq},
which form the following system of differential equations
\begin{equation}
\label{2phase-eq1}
\begin{aligned}
& \dot x = Ax + \frac{b}{2}\sin(\theta_{\Delta}),
\\
& \dot\theta_{\Delta} = \omega_{\Delta} - Lc^*x - \frac{Lh}{2}\sin(\theta_{\Delta}),
\\
& \theta_{\Delta}(t) = \theta_1(t) - \theta_2(t),
  \quad
  \omega_{\Delta} = \omega_1 - \omega_{free}
\end{aligned}
\end{equation}
For a lead-lag filter, described by the transfer function
$H(s) = \frac{1+s \tau_2}{1+s(\tau_1 + \tau_2)}$,
system \eqref{2phase-eq1} takes the form
\begin{equation}
\label{2phase-eq2}
\begin{aligned}
& \dot x =
  \frac{-1}{\tau_1 + \tau_2}x
  + (1-\frac{\tau_2}{\tau_1 + \tau_2})\frac{1}{2}\sin(\theta_{\Delta}),
\\
& \dot\theta_{\Delta} =
  \omega_{\Delta}
  - L\frac{1}{\tau_1 + \tau_2}x \
  - \frac{\tau_2}{\tau_1 + \tau_2}\frac{L}{2}\sin(\theta_{\Delta}).
\\
\end{aligned}
\end{equation}
The equilibrium points of \eqref{2phase-eq2} are defined by the following relations:
\begin{equation}
\label{equilibriums}
\begin{aligned}
& x_{eq} = \frac{\tau_1}{2}\sin(\theta_{\Delta}),
\
\sin(\theta_{eq}) = 2\frac{\omega_{\Delta}}{L} .
\end{aligned}
\end{equation}
For $\tau_1 = 0.0448$,
 $L = 500$, and $\omega_{\Delta} = 178.9$
we get
\begin{equation}
\label{equilibriums}
\begin{aligned}
& x_{eq} = 0.016,
\\
& \theta_{eq} = (-1)^k 0.7975 + \pi k, \quad k \in \mathbb{N}.
\\
\end{aligned}
\end{equation}

\begin{figure}[H]
  \includegraphics[width=0.85\linewidth]{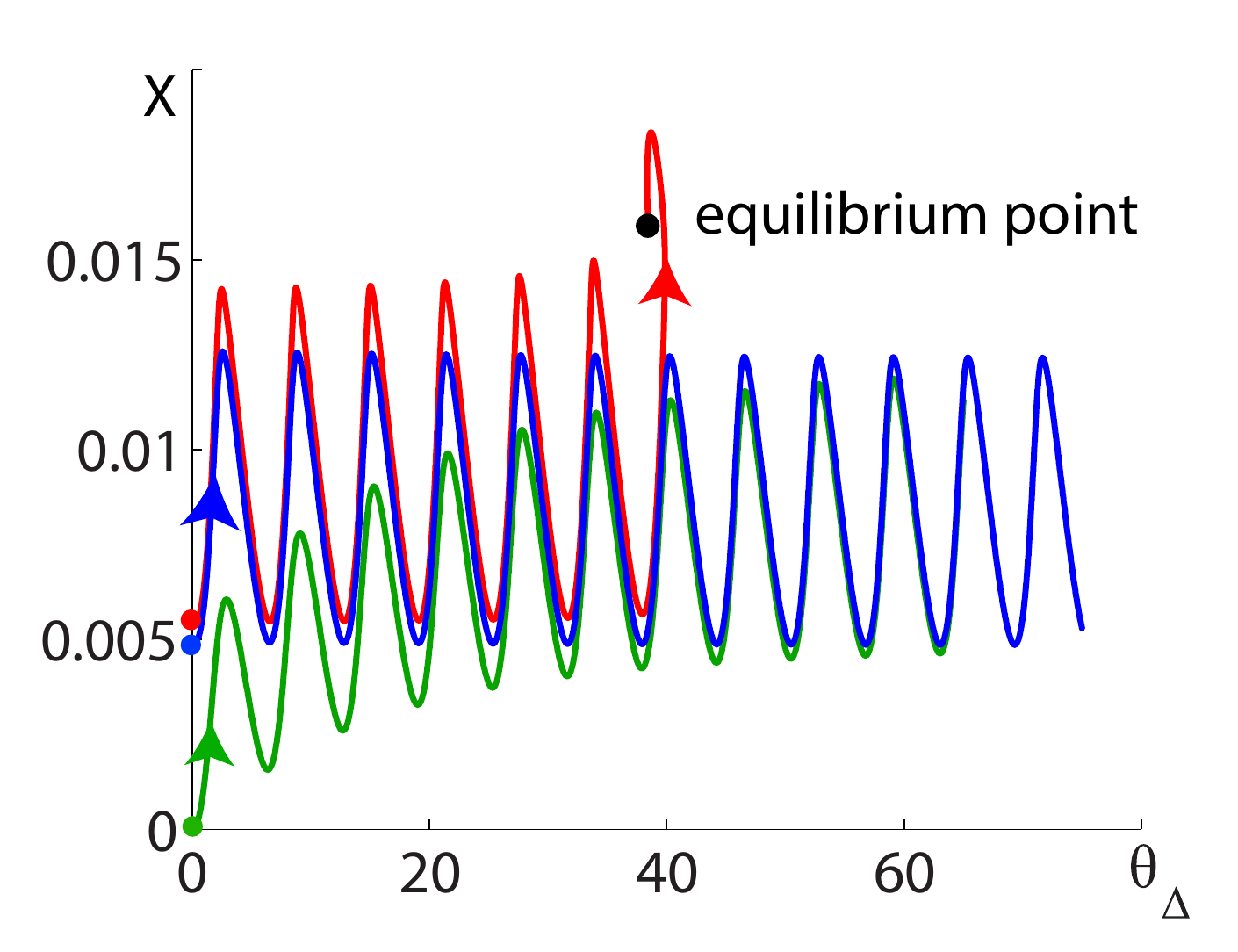}
  \caption{Phase portrait of the classical PLL with stable\
           and unstable periodic trajectories}
  \label{pll_hidden_phase_portret}
\end{figure}

Consider now a phase portrait
(where the system's evolving state over time traces a trajectory $(x(t),\theta_\Delta(t))$),
corresponding to signal's phase model (see Fig.~\ref{pll_hidden_phase_portret}).
\begin{figure}[h]
  \includegraphics[width=0.85\linewidth]{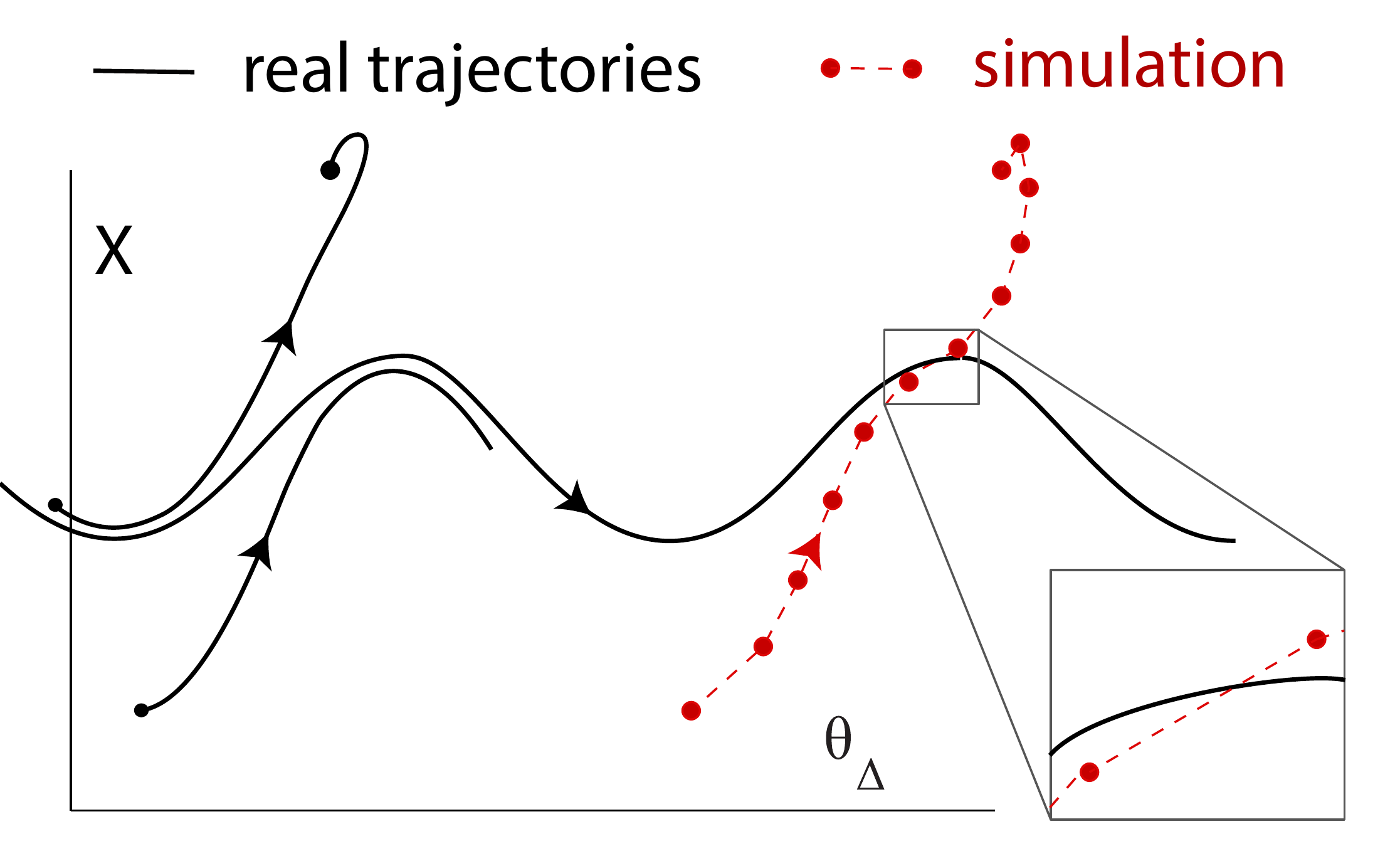}
  \caption{Phase portrait of the classical PLL with stable\
   and unstable periodic trajectories}
  \label{pll_hidden_phase_portret3}
\end{figure}
The solid blue line in Fig.~\ref{pll_hidden_phase_portret} corresponds
to the trajectory with the loop filter initial state $x(0) = 0.005$
and the VCO phase shift $0$ rad.
This line tends to the periodic trajectory,
therefore it will not acquire lock.
All the trajectories under the blue line
(see, e.g., a green trajectory with the initial state $x(0) = 0$)
also tend to the same periodic trajectory.

The solid red line corresponds to the trajectory
with the loop filter initial state $0.00555$
and the VCO initial phase $0$.
This trajectory lies above the unstable periodic trajectory and tends
to  a stable equilibrium.
In this case PLL acquires lock.

All the trajectories between stable and unstable periodic trajectories
tend to the stable one (see, e.g., a solid green line).
Therefore, if the gap between stable and unstable periodic trajectories
is smaller than the discretization step,
the numerical procedure may slip through the stable trajectory.
The case corresponds to the close coexisting attractors
and the bifurcation of birth of semistable trajectory
\cite{LeonovK-2013-IJBC,KuznetsovLYY-2014-IFAC}.
In this case numerical methods are limited
by the errors on account of the linear multistep integration methods (see \cite{biggio2013reliable,biggio2014accurate}).
As noted in \cite{Brambilla-2003-frequency},
low-order methods introduce a relatively large warping error that, in some
cases, could lead to corrupted solutions (i.e., solutions that are wrong
even from a qualitative point of view).
This example demonstrate also the difficulties of numerical search
of so-called hidden oscillations, whose basin
of attraction does not overlap with the neighborhood
of an equilibrium point, and thus may be difficult to
find numerically\footnote{
An oscillation in a dynamical system can be easily localized
numerically if the initial data from its open neighborhood lead to
long-time behavior that approaches the oscillation.
From a computational point of view, on account of
the simplicity of finding the basin of attraction in the phase space,
it is natural to suggest the following classification of attractors
\cite{KuznetsovLV-2010-IFAC,LeonovKV-2011-PLA,LeonovKV-2012-PhysD,LeonovK-2013-IJBC,LeonovKM-2015-EPJST}:
{\it An attractor is called a \emph{hidden attractor} if its
 basin of attraction does not intersect
 small neighborhoods of equilibria,
 otherwise it is called a \emph{self-excited attractor}.
}

For a \emph{self-excited attractor} its basin of attraction
is connected with an unstable equilibrium. Therefore, self-excited attractors
can be localized numerically by the \emph{standard computational procedure}
in which after a transient process a trajectory,
started from a point of unstable manifold in
a neighborhood of unstable equilibrium,
is attracted to the state of oscillation and traces it.
Thus self-excited attractors can be easily visualized.

In contrast, for a hidden attractor its basin of attraction
is not connected with unstable equilibria.
For example, hidden attractors can be attractors in
the systems with no equilibria
or with only one stable equilibrium
(a special case of multistable systems and
coexistence of attractors).
Recent examples of hidden attractors can be found in
\emph{The European Physical Journal Special Topics: Multistability: Uncovering Hidden Attractors}, 2015
(see \cite{Shahzad20151637,Brezetskyi20151459,Jafari20151469,Zhusubaliyev20151519,Saha20151563,Semenov20151553,Feng20151619,Li20151493,Feng20151593,Sprott20151409,Pham20151507,Vaidyanathan20151575}).
}.
In this case the observation of one or
another stable solution may depend on the initial data and
integration step.

%
%

\section{Conclusion}

The considered example is a motivation for the use of
rigorous analytical methods
for the analysis of nonlinear PLL models.
Rigorous study of the above effect can be done by Andronov's point transformation method
and phase plane analysis.
Corresponding bifurcation diagram were given in \cite{Belyustina-1970-eng} (see, also \cite{LeonovK-2013-IJBC,LeonovK-2014}).


For Costas loop models, similar effect is shown in
\cite{KuznetsovKLNYY-2014-ICUMT-QPSK,KuznetsovKLSYY-2014-ICUMT-BPSK,KudryashovaKKLSYY-2014-ICINCO}.

%


\bibliographystyle{IEEEtran}

\end{document}